%% file: ifip-kumodd.tex
\documentclass[titlepage]{report}

\usepackage[export]{adjustbox}
\usepackage[titletoc]{appendix}
\usepackage{color}
\usepackage{float}
\usepackage[letterpaper, portrait, margin=1in]{geometry}
\usepackage{graphicx}
\usepackage[utf8]{inputenc}
\usepackage{listings}
\usepackage{placeins}
\usepackage{soul}
\usepackage{todonotes}
\usepackage[normalem]{ulem}
\usepackage{url}
\usepackage{xspace}

\usepackage{setspace}
\usepackage{caption}
\usepackage{tikz}
\usetikzlibrary{arrows,positioning,shapes.geometric}

\usepackage{etoolbox}
\patchcmd{\thebibliography}{\chapter*}{\section*}{}{}

\definecolor{codegreen}{rgb}{0,0.6,0}
\definecolor{codegray}{rgb}{0.5,0.5,0.5}
\definecolor{codepurple}{rgb}{0.58,0,0.82}
\definecolor{backcolour}{rgb}{0.95,0.95,0.92}

\lstdefinestyle{mystyle}{
	backgroundcolor=\color{backcolour},
	commentstyle=\color{codegreen},
	keywordstyle=\color{magenta},
	numberstyle=\tiny\color{codegray},
	stringstyle=\color{codepurple},
	basicstyle=\footnotesize\ttfamily,
	breakatwhitespace=false,
	breaklines=true,
	captionpos=b,
	keepspaces=true,
	showspaces=false,
	showstringspaces=false,
	showtabs=false,
	tabsize=2
}
\graphicspath{ {images/} }

\newcommand*{\kumodd}{\texttt{kumodd}\xspace}
\newcommand*{\Kumodd}{\textit{Kumodd}\xspace}
\newcommand*{\gdrive}{\emph{Google Drive}\xspace}
\newcommand*{\gdocs}{\emph{Google Docs}\xspace}
\newcommand*{\dbox}{\emph{Dropbox}\xspace}
\newcommand*{\onedrive}{\emph{Microsoft OneDrive}\xspace}
\newcommand*{\boxnet}{\emph{Box}\xspace}

\renewcommand*{\thefootnote}{\fnsymbol{footnote}}
\begin{document}

\title{Forensic Acquisition of Cloud Drives\footnote{Please cite as:
\ Vassil Roussev, Andres Barreto, and Irfan Ahmed, \emph{Forensic Acquisition of Cloud Drives}, In \emph{Advances in Digital Forensics XII}, Gilbert Peterson and Sujeet Shenoi (eds.), Springer, 2016.}
}
\author{Vassil Roussev\footnote{Corresponding author, \texttt{vassil@roussev.net}.}, Andres Barreto, Irfan Ahmed\\
Greater New Orleans Center for Information Assurance\\
University of New Orleans\\
\ \\
\\
}
\maketitle
\renewcommand*{\thefootnote}{\arabic{footnote}}
\setcounter{footnote}{0}
\setlength{\footnotesep}{\baselineskip}

\pagenumbering{roman}
\setcounter{page}{2}

\section*{Abstract}

Cloud computing and cloud storage services, in particular, pose a new challenge to digital forensic investigations.
Currently, evidence acquisition for such services still follows the traditional method of collecting artifacts on a \emph{client} device.
This approach requires labor-intensive reverse engineering efforts, and ultimately result in an acquisition that is \emph{inherently} incomplete.
Specifically, it makes the incorrect assumption that \emph{all} storage content for an account is fully replicated on the client;
further, there are no means to acquire historical data in the form of document revisions, nor is there a way to acquire cloud-native artifacts, such as \gdocs.

In this work, we introduce the concept of \emph{API-based evidence acquisition} for cloud services, which addresses these concerns by utilizing the officially supported API of the service.
To demonstrate the utility of this approach, we present a proof-of-concept acquisition tool, \kumodd, which can acquire evidence from four major cloud drive providers: \emph{Google Drive}, \emph{Microsoft OneDrive}, \emph{Dropbox}, and \emph{Box}.
The implementation provides both command-line and web user interfaces, and can be readily incorporated into established forensic processes.

\paragraph{Keywords: }
Cloud forensics, cloud storage forensics, kumodd, API-based evidence acquisition, Google Drive, Dropbox, OneDrive, Box.

\renewcommand{\thepage}{\arabic{page}}
\setcounter{page}{1}

\input{sections/01-introduction}
\input{sections/02-related}

\input{sections/03-need}

\input{sections/04-kumodd}

\input{sections/05-discussion}
\input{sections/06-conclusion}

\bibliographystyle{IEEEtran}
\bibliography{ifip-kumodd}

\end{document}

%% file: sections/01-introduction.tex
\section*{Introduction}

Cloud computing is the emerging primary model for delivering information technology (IT) services to Internet-connected devices.
It abstracts away the physical compute and communication infrastructure, and allows customers to effectively rent, instead of own and maintain, as much compute capacity as needed.
As per NIST's definition \cite{nist-cloud}, there are five essential characteristics--\emph{on-demand self service}, \emph{broad network access}, \emph{resource pooling}, \emph{rapid elasticity}, and \emph{measured service}--that distinguish the cloud service model from previous ones.

The underpinning technology development that has made the cloud possible is the massive adoption of virtualization on commodity hardware systems.
Ultimately, it allows for a large pool of resources, such as a data center, to be provisioned and load-balanced at a fine granularity, and for the computations of different users and uses to be strongly isolated.

The first public cloud services--\emph{Amazon Web Services} (AWS)--were introduced by Amazon in 2006.
As of 2015, according to RightScale’s \emph{State of the Cloud Report} \cite{rightscale-2015}, cloud adoption has become ubiquitous:
93\% of businesses are at least experimenting with cloud deployments, with 82\% adopting a hybrid strategy, which combines the use of multiple providers (usually in a public-private configuration). Nonetheless, much of the technology transition is still ahead as 68\% of enterprises have less than 20\% of their application portfolio running in a cloud setup. Similarly, \emph{Gartner} \cite{gartner-2014} predicts another 2-5 years will be needed before cloud computing reaches the ``plateau of productivity''\cite{gartner-hype} marking the period of mass mainstream adoption and widespread productivity gains.

Unsurprisingly, cloud forensics is still in its infancy; there are few practical solutions for the acquisition and analysis of cloud evidence and most of them are minor adaptations of existing methods and tools.
Indeed, NIST--the main standardization body in the US--is still working to build consensus on what the \emph{challenges} are with respect to performing forensics of cloud data;
in \cite{nist-cloud}, the authors have enumerated 65 separate ones.

In this work, we are concerned with one specific problem--the acquisition of data from cloud storage services.
These have emerged as one of the most popular services with consumers as many providers, such as \dbox, \boxnet, \gdrive, and \onedrive, provide between 2 and 15GB of cloud storage for free.
Cloud storage is also widely used on mobile devices to share data across applications (which are otherwise isolated from each other).
Therefore, having a robust evidence acquisition method is a necessity \emph{today};
further, due to the wide variety of these services, and the rapid introduction of new ones, the tool and methodology should be adaptable and extensible.

In traditional forensic models, the investigator works with physical evidence carriers, such as storage media or integrated compute devices (e.g., smartphones).
Thus, it is easy to identify the computer performing the computations and the media that store (traces of) the processing, and to physically collect, preserve and analyze the relevant information content.
Because of this, research has focused on discovering and acquiring every little piece of log and timestamp information, and extracting every last bit of discarded data that applications and the OS have left behind.

Conceptually, cloud computing breaks this model in two major ways.
First, resources--CPU cycles, RAM, storage, etc.--are pooled (e.g., RAID storage) and then allocated at a fine granularity.
This results in physical media usually containing data owned by many users;
as well, data relevant to a single case can end up spread among numerous storage media and (potentially) among different providers.
Applying the conventional model creates a long list of procedural, legal, and technical problems that are unlikely to have an efficient solution in the general case.
Second, both computations and storage contain a much more ephemeral record as virtual machine (VM) instances are created and destroyed with regularity and working storage is routinely sanitized.

As we discuss in the next section, current work on cloud storage forensics has treated the problem as just another instance of application forensics.
It applies basic differential analysis techniques to gain an understanding of the artifacts present on client devices by taking before and after snapshots of the target compute system, and deducing relevant cause and effect relationships.
During an actual investigation, the analyst would be interpreting the state of the system based on these known relationships.

\vspace{6pt}
Unfortunately, there are several serious problems with this application of existing client-side methods:

\begin{itemize}
\item \emph{Completeness.}
The reliance in client-side data can leave out critical case data.
The most basic example is the presense of selective replication of cloud drive data, which means that the client device may simply not have a copy of all the stored data locally.
As use grows--\gdrive already offers up to 30TB--this will be increasingly the typical case.

\item \emph{Correctness \& reproducibility.}
Conceptually, it is not feasible to reverse engineer \emph{all} aspects of an application's functionality if the source is not available, which immediately calls into question the correctness of the analysis.
Further, cloud storage applications on the client get updated frequently with new features introduced on a regular basis.
This places a burden on forensics to keep up the reversing engineering efforts and it becomes harder to maintain the reproducibility of the analysis.

\item \emph{Cost \& scalability.}
As a continuation of the prior point, manual client-side analysis is burdensome and simply does not scale with the rapid growth of the variety of services and their versions.

\end{itemize}

We present an alternative approach for the acquisition of evidence data from cloud storage providers that uses the official APIs provided by the services.
Such an approach has the immediate advantage of taking all the reverse-engineering work out of the picture, and the following \emph{conceptual} advantages:

\begin{itemize}
\item APIs are well-documented, official interfaces through which cloud applications on the client communicate with the service;
they tend to change slowly and changes are clearly marked--only new features need to be incrementally incorporated into the acquisition tool.

\item It is easy to demonstrate completeness and reproducibility using the API specification.

\item Web APIs tend to follow patterns, which makes it possible to adapt existing code to a new (similar) service with modest effort.
It is often feasible to write an acquisition tool for a completely new service from scratch in a few hours.
\end{itemize}

To demonstrate the feasibility of our approach, and to gain first-hand experience with the process, we have developed a proof-of-concept prototype called
\emph{kumodd}\footnote{The tool name is derived from the Japanese word for cloud (\emph{kumo}) and the venerable \emph{dd} Unix utility.}, which can perform complete (or partial) acquisition of a cloud storage account's data.
It works with four popular services--\emph{Dropbox}, \emph{Box}, \emph{Google Drive}, and Microsoft's \emph{OneDrive}--and supports the acquisition of revisions and cloud-only documents.
The prototype is written in \emph{Python} and offers both a command line and web-based user interfaces.

%% file: sections/02-related.tex
\section*{Related Work}

In this section, we briefly summarize essential cloud terminology and discuss representative related work.

\subsection*{Background: cloud computing}
The reference definition of cloud computing provided by the National Institute of Standards and Technology (NIST) states that
``cloud computing is a model for enabling ubiquitous, convenient, on-demand network access to a shared pool of configurable computing resources (e.g., networks, servers, storage, applications, and services) that can be rapidly provisioned and released with minimal management effort or service provider interaction" \cite{nist-cloud}.
With respect to \emph{public} cloud services--the most common case--this means that the physical hardware on which the computation takes place is owned and maintained the provider, as is at least part of the deployed software stack.
Generally, customers have the option to pay per unit of CPU/storage/network use, although other business arrangements are also possible.

\begin{figure}[H]
	\centering
	\includegraphics[width=1\textwidth]{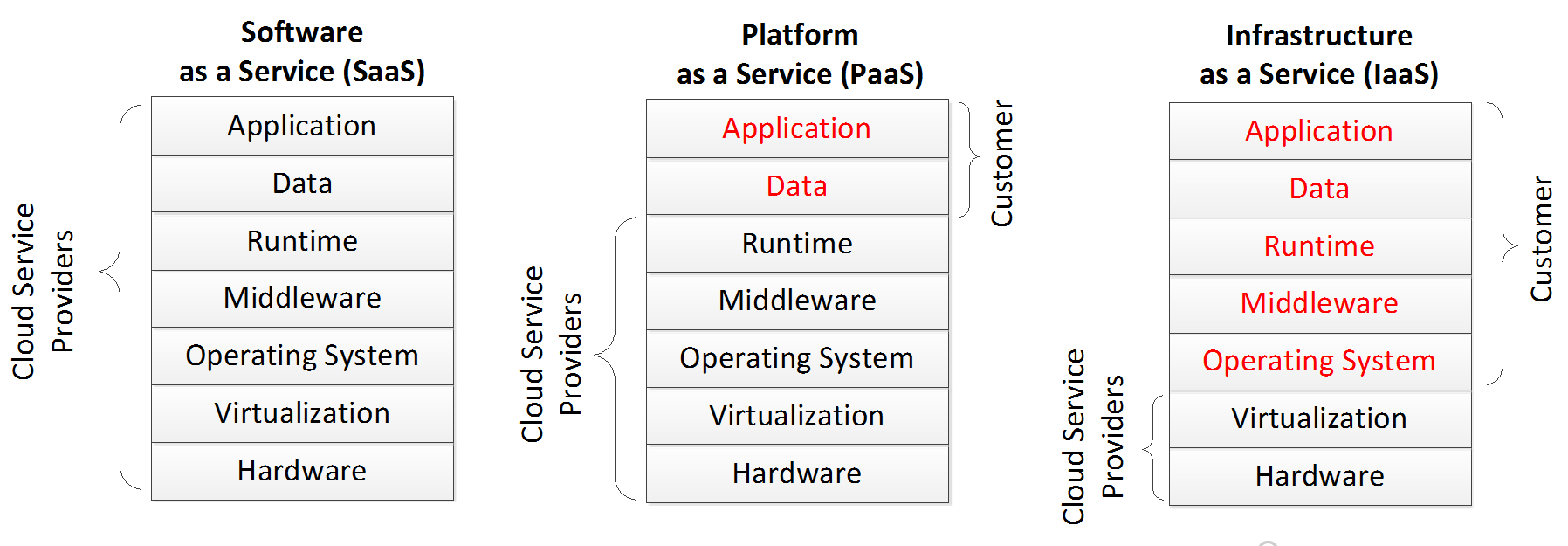}
	\caption{Cloud service models and ownership of layers (public cloud)}
	\label{fig:cloud-stack}
\end{figure}

Cloud computing services are commonly classified into one of three canonical models—\emph{software as a service} (SaaS), \emph{platform as a service} (PaaS), and \emph{infrastructure as a service} (IaaS).
In reality the distinctions are often less clear cut, and practical IT cloud solutions--and potential investigative targets--can incorporate elements of all of these.
As illustrated on Figure \ref{fig:cloud-stack}, it is useful to decompose cloud-computing environments into a stack of layers (from lower to higher): \emph{hardware}, \emph{virtualization} consisting of hypervisor allowing to install virtual machines, \emph{operating system} installed on each virtual machine, \emph{middleware} and \emph{runtime} environment, and \emph{application} and \emph{data}.

Depending on the deployment scenario, different layers could be managed by different parties.
In a private deployment, the entire stack is hosted by the owner and the overall forensic picture is nearly identical to the case of investigating a non-cloud IT target.
Data ownership is clear, as is the legal and procedural path to obtain it; indeed, the very use of the term cloud is mostly immaterial to forensics.
In a public deployment, the SaaS/PaaS/IaaS classification becomes important as it defines the ownership and management responsibilities over data and service (Figure~\ref{fig:cloud-stack}).
In hybrid deployments, layer ownership can be split between the customer and the provider and/or across \emph{multiple} providers.
Further, this relationship can change over time as, for example, the customer may handle the base load on owned infrastructure, but burst to the public cloud to handle peak demand, or system failures.

The main point here is that the potential targets of cloud forensics can vary widely and, in our view, the most productive approach to developing practical solutions is to start with specific (but common) cases and, over time, attempt to incorporate an expanding range.
The focus of our work is the forensics of cloud drive services, starting with the acquisition process.

\subsection*{Cloud drive forensics}

The concept of a ``cloud drive'' is closely related to network filesystem shares and is hardly distiguishable from different versions of the \emph{i-drive} (Internet drive) that became popular the late 1990s.
The main difference is that of scale--today, there are many more providers and the WAN infrastructure has much higher bandwidth capacity, which makes real-time file synchronization much more practical.
There are many more providers, most of which build their services in top of third party IaaS offerings (such as AWS).
Over the last few years, a number of forensic researchers have worked on cloud drives.

Chung et al. \cite{article-chung} analyzed four cloud storage services (\emph{Amazon S3}, \gdocs, \dbox, and \emph{Evernote}) in search of traces left by them on the client system that can be used in criminal cases.
They reported that the analyzed services may create different artifacts depending on specific features of the services, and proposed a process model for forensic investigation of cloud storage services which is based in the collection and analysis of artifacts of the analyzed cloud storage services from client systems.
The procedure includes gathering volatile data from a Mac or Windows system (if available), and then retrieving data from the Internet history, log files, and directories.
In mobile devices they rooted an Android phone to gather data and for iPhone they used iTunes information like backup iTunes files.
The objective was to check for traces of a cloud storage service exist in the collected data.

In \cite{hale13-amazon-drive}, Hale analyzes the \emph{Amazon Cloud Drive} and discusses the digital artifacts left behind after an Amazon Cloud Drive account has been accessed or manipulated from a computer.
There are two possibilities to manipulate an Amazon Cloud Drive Account: one is via the web application accessible using a web browser and the other is a client application provided by Amazon and can be installed in the system.
After analyzing the two methods he found artifacts of the interface on web browser history, and cache files.
He also found application artifacts in the Windows registry, application installation files on default location, and an SQLite database used to keep track of pending upload/download tasks.

Quick et al. \cite{quick13-dropbox} analyzed \dbox and discusses the artifacts left behind after a \dbox account has been accessed, or manipulated.
Using hash analysis and keyword searches they try to determine if the client software provided by \dbox has been used.
This involves extracting the account username from browser history (Mozilla Firefox, Google Chrome, and Microsoft Internet Explorer), and the use of the \dbox through several avenues such as directory listings, prefetch files, link files, thumbnails, registry, browser history, and memory captures.
In follow-up work, Quick et al. \cite{quick14-gdrive} use a similar conceptual approach to analyze the client-side operation and artifacts of \gdrive and provide a starting point for investigators.

Martini and Choo \cite{martini13-owncloud} have researched the operation of \emph{ownCloud}, which is a self-hosted file synchronization and sharing solution.
As such, it occupies a slightly different niche as it is much more likely for the client and server sides to be under the control of the same person/organization.
They were able to recover artifacts including sync and file management metadata (logging, database and configuration data), cached files describing the files the user has stored on the client device and uploaded to the cloud environment or vise versa, and browser artifacts.

\subsection*{Summary}
In brief, previous work on cloud storage forensics has been focused on adapting the traditional application forensics approach to finding client-side artifacts.
This involves blackbox differential analysis, where before and after images are created and compared to deduce the essential functions of the application.

Clearly, the effectiveness of the above approach depends on the comprehensiveness of the tests performed on the target system;
ultimately, it is nearly impossible to enumerate all eventualities that may affect the state of the application.
The process is a labor-intensive reverse engineering effort, which requires substantial investment of human resources.
Yet, as discussed in the following section, the biggest limitation of client-side forensics is that it \emph{cannot} guarantee full acquisition of the cloud drive data.

%% file: sections/03-need.tex
\section*{New Approach: API-based Acquisition}
\subsection*{The lapses of client-side acquisition}

\begin{figure}[H]
	\centering
	\includegraphics[width=\textwidth,trim={0 7.75in 0 0.5in}]{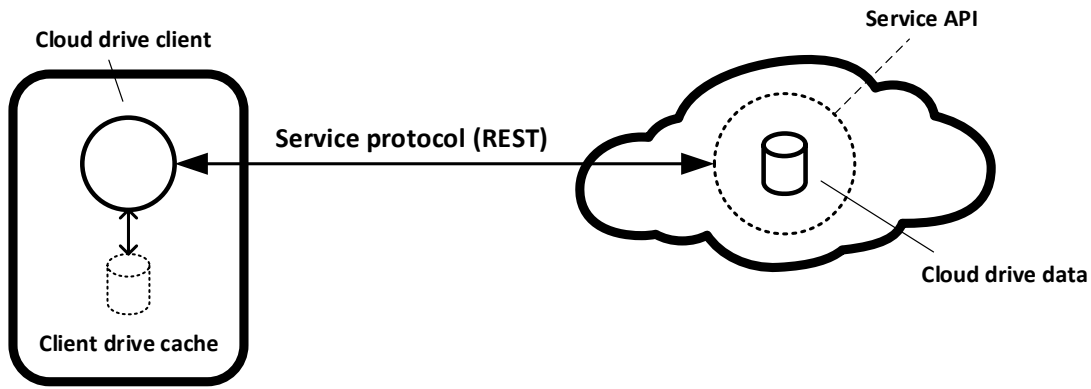}
	\caption{Cloud drive service: architectural diagram}
	\label{fig:cloud-drive-arc}
\end{figure}

The easiest way to understand the problems of client-side acquisition is to recognize that, for cloud data, it is an acquisition-by-proxy process.
In other words, although it resembles traditional acquisition from physical media, this method does not target the authoritative source of the data--the cloud.
As illustrated by Figure~\ref{fig:cloud-drive-arc}, client content is properly viewed as a cached copy of cloud-hosted data; this simple fact has crucial implications for forensic acquisition.

\emph{Partial replication.}
The most obvious problem is that there is no guarantee that \emph{any} of the clients attached to an account will have a a complete copy of the (cloud) drive's content.
As a point of reference, \gdrive currently offers up to 30TB of online storage (at \$10/TB per month), whereas Amazon offers \emph{unlimited} storage at \$60/year.
As data accumulates online, it quickly becomes impractical to keep full replicas on all devices; indeed, with current trends, it is likely that \emph{most} users will have \emph{no} device with a complete copy of the data.
Forensically speaking, we need direct access to the cloud drive's metadata to ascertain its contents; the alternative, blindly relying on client cache, can only be labeled as incomplete acquisition with unknown gaps.

\emph{Revisions.} Most drive services provide some form of revision history; the lookback period varies between 30 days and unlimited revision history (depending on the service and subscription).
This is a new source of valuable forensic information that has few analogs in traditional forensic targets (e.g., Volume Shadow Copy service on Windows) and investigators are not yet used to looking for it.
Revisions reside in the cloud and clients rarely have anything but the most recent version in their cache; a client-side acquisition will clearly miss prior revisions and will not even know that it is missing.

\emph{Cloud-native artifacts.}
Courtesy of the wholesale movement to web-based applications, forensics needs to learn how to deal with a new problem--digital artifacts that have \emph{no} serialized representation in the local filesystem.
For example, \gdocs documents are stored locally as a link to the document, which can only be edited via a web app.
Acquiring an opaque link, by itself, is borderline useless--it is the content of the document that is of primary interest.
It is typically possible to obtain a usable snapshot of the web app artifact (e.g., in PDF) but that can only be accomplished by requesting it from the service directly;
again, acquisition-by-proxy cannot accomplish that.

In sum, even our brief examination here shows that the client-side approach to drive acquisition has major \emph{conceptual} flaws that are beyond remediation;
we need a different method that obtains the data directly from the cloud service.

\subsection*{Learning to love the API}
Fortunately, cloud services provide a front door--an API--to directly acquire the content of a cloud drive.
In broad terms, a cloud drive provides a storage service similar to that of a local filesystem--it enables the creation and organization of user files.
Therefore, its API loosely resembles that of the filesystem API provided by the local operating system.
Before we continue with the technical details of our proof-of-concept tool, we believe it is necessary to make the case that the use of the API is forensically sound.

The main issue to address head-on is the fact that using an API-based approach results in \emph{logical} evidence acquisition, not physical.
Traditionally, it has been held as an article of faith that obtaining data at the lowest possible level of abstraction results in the most reliable evidence.
The main rationale is that the logical view of the data may not be forensically complete as data marked as deleted will not be shown;
also, a technically sophisticated adversary may be able to hide data from the logical view.
Until a few years ago, this view would have been reasonably  justified.

However, it is important to periodically examine our accepted wisdom to account for new technology developments.
It is outside the scope of this paper to make a more general argument, but we would note that solid-stated drives (SSDs) and even newer generations of high-capacity HDDs much more resemble autonomous storage computers than the limited peripherals of 10 years ago.
Some of them contain ARM processors and execute complex load-balancing and wear-leveling algorithms, which include background data relocation.
Although they support, for example, block-level access, the results do not directly map to a physical layout of the data; this makes the acquired image logical, not physical.
To obtain (and make sense of) a trully low-level representation of the data would increasingly requires \emph{hardware} blackbox reverse engineering.
More than likely, this will lead to the wider acceptance of de facto logical acquisition as forensically sound.

For cloud forensics, the case for adopting API-mediated acquisition \emph{now} is simple and unambiguous.
As per Figure~\ref{fig:cloud-drive-arc}, the client component of the cloud drive (which manages the local cache) utilizes the exact same interface to perform its operations.
Thus, the service API is the lowest available level of abstraction and is, therefore, appropriate for forensic processing.
Further, metadata for individual files often includes cryptographic hashes of the content, which enables strong integrity guarantee during acquisition.

The service API (and corresponding client SDKs for different languages) are officially supported by the provider and have well-defined semantics and detailed documentation;
this allows for formal and precise approach to forensic tool development and testing.
In contrast, blackbox reverse engineering can never achieve provable perfection.
Similarly, acquisition completeness guarantee can only be achieved via the API--the client cache contains an unknown fraction of the content.

Finally, software development is almost always easier and cheaper than reverse engineering followed by software development.
The core of our PoC prototype is less than 1,600 lines of Python code (excluding the web UI) for four different services.
An experienced developer could easily add a good-quality driver for a new (similar) service in a day, or two, including test code.
The code only needs to be updated infrequently as providers strive to provide continuity and backward compatibility;
any relevant additions to the API can easily be identified and incrementally adopted.

%% file: sections/04-kumodd.tex
\section*{Design \& Implementation of \texttt{kumodd}}
\begin{figure}[H]
	\centering
    \begin{tikzpicture}[>=latex']
        \tikzset{
            block/.style=  {draw, rectangle, align=center,minimum width=2cm,minimum height=1cm},
            rblock/.style= {draw, shape=rectangle,rounded corners=1.5em,align=center,minimum width=2cm,minimum height=1cm},
            trap/.style= {
                draw,
                trapezium,
                trapezium left angle=60,
                trapezium right angle=120,
                minimum width=1.5cm,
                align=center,
                minimum height=1cm
            },
        }
        \node [rblock]  (cloud) {Cloud \\ drive};
        \node [block, right =1cm of cloud] (discovery) {Content \\ discovery};
        \node [block, right =1cm of discovery] (selection) {Target \\ selection};
        \node [block, right =1cm of selection] (acquisition) {Target \\ acquisition};
        \node [trap, right =1cm of acquisition] (evidence) {Acquired \\ evidence};
        \path[draw,->] (cloud) edge (discovery)
                    (discovery) edge (selection)
                    (selection) edge (acquisition)
                    (acquisition) edge (evidence)
                    ;
    \end{tikzpicture}
	\caption{Acquisition phases}
	\label{fig:acquisition-phases}
\end{figure}
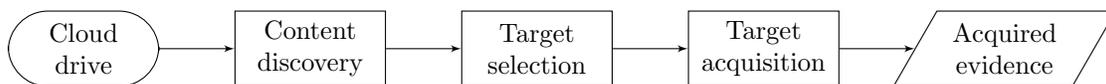

Conceptually, acquisition consists of three core phases--content discovery, target selection, and target acquisition (Figure~\ref{fig:acquisition-phases}).
During content discovery, the acquisition tool queries the target and obtains a list of artifacts (files) along with their metadata.
In a baseline implementation this can be reduced to enumerating all available files; in a more advanced one, the tool can take advantage of search capability provided by the API (e.g., \gdrive).
During the selection process, the list of targeted artifacts can be filtered down by automated means, or by involving the user.
The result is a (potentially prioritized) list of targets that is passed onto the tool to acquire.

Traditional approaches largely short-circuit this process by attempting to blindly acquire \emph{all} available data.
However, this ``acquire-first-filter-later'' approach is not remotely sustainable for cloud targets--the overall amount of data can be enormous and the available bandwidth could be up to two orders of magnitude lower than local storage.

The goal of \kumodd is to be a minimalistic tool for research and experimentation that can also provide a basic practical solution for real cases;
we have sought to make it as simple as possible to integrate it with the existing toolset.
Its basic operation is to acquire (a subset of) the content of a cloud drive and place it an appropriately structured local filesystem tree.

\subsection*{Architecture}

\begin{figure}[H]
	\centering
	\includegraphics[width=0.8\textwidth]{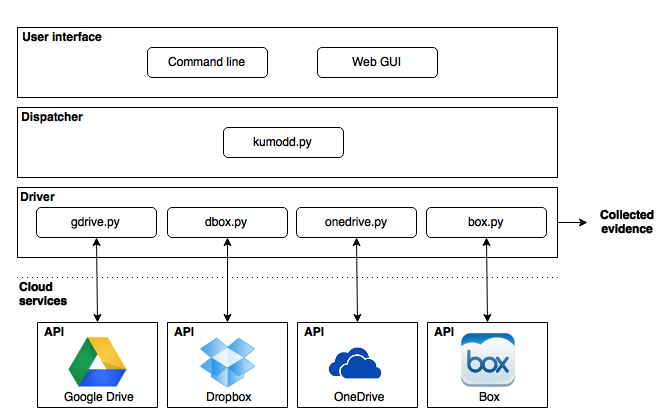}
	\caption{\Kumodd architectural diagram}
	\label{fig:kumodd-design}
\end{figure}

\Kumodd is split into several modules and in three logical layers--dispatcher, drivers, and user interface (Figure~\ref{fig:kumodd-design}).
The dispatcher (\texttt{kumodd.py}) is the central component which receives parsed user requests, relays them to the appropriate driver, and sends back the result.
The drivers--one for each service--implement the provider-specific protocol via the web API.
The tool provides two interface--a command-line one (CLI) and a web-based GUI.

\subsection*{Command-line tool use}
The general format of the \kumodd commands is:

\begin{lstlisting}
python kumodd.py -s [service] [action] [filter]
\end{lstlisting}

The \texttt{[service]} parameter specifies the target service.
Currently, the supported options are \texttt{gdrive}, \texttt{dropbox}, \texttt{onedrive}, and \texttt{box}, which correspond to \gdrive, \dbox, \onedrive, and \boxnet, respectively.

The \texttt{[action]} argument instructs \kumodd on what to do with the target drive:
\texttt{-l\ } \emph{list} stored files (as a plain text table);
\texttt{-d\ } \emph{download} files (subject to the \texttt{[filter]} specification);
and \texttt{-csv <file>\ } download the files specified by the file (in CSV format).
The \texttt{-p <path>\ } option can be used to explicitly specify the path to which the files should be downloaded (and override the default, which is relative to the current working directory).

The \texttt{[filter]} parameter specifies the subset of files to be listed/downloaded based on file type:
\texttt{all}--all files present; \texttt{doc}--all Microsoft Office/Open Office document files (\texttt{.doc/.docx/.odf}); \texttt{xls}--spreadsheet files; \texttt{ppt}--presentations files; \texttt{text}--text/source code;
\texttt{pdf}--PDF files.
In addition, some general groups of files can also be specified: \texttt{officedocs}--all document, spreadsheet and presentation files; \texttt{image} --all images; \texttt{audio}--all audio files; and \texttt{video}--all video files.

\noindent\emph{Examples}
\vspace{-9pt}
\begin{itemize}
\item[--] List all files stored in a \dbox account:
\end{itemize}
\vspace{-9pt}
\begin{lstlisting}
python kumodd.py -s dbox -l all
\end{lstlisting}

\begin{itemize}
\item[--] List images stored in a \boxnet account
\end{itemize}
\vspace{-9pt}
\begin{lstlisting}
python kumodd.py -s box -l image
\end{lstlisting}

\begin{itemize}
\item[--] Download the PDF files stored in a \onedrive account to the Desktop folder
\end{itemize}
\vspace{-9pt}
\begin{lstlisting}
python kumodd.py -s onedrive -d all -l -p /home/user/Desktop/
\end{lstlisting}

\begin{itemize}
\item[--] Download the files listed in \emph{gdrive\_list.csv} from \gdrive
\end{itemize}
\vspace{-9pt}
\begin{lstlisting}
python kumodd.py -s gdrive -csv /home/user/Desktop/gdrive_list.csv
\end{lstlisting}

\subsubsection*{User authentication}
All four of the services use the OAuth2 (\url{http://oauth.net/2/}) protocol to authenticate the user and to authorize access to the account.
When \kumodd is used for the first time to connect to a cloud service, the respective driver will initiate the authorization process which requires the user to authenticate with the appropriate credentials (username/password).
The tool provides the user with a URL that needs to be opened in a web browser, where the standard authentication interface for the service will request the relevant username and password. We illustrate the process using \gdrive as an example:

\begin{lstlisting}[title=Authentication step 1: connect to \gdrive]
kumo@ubuntu:~/kumodd$ python kumodd.py -s gdrive -d all
Your browser has been opened to visit:
https://accounts.google.com/o/oauth2/auth?scope=https%3A%2F%2Fwww.googleapis.com...
...
\end{lstlisting}

\begin{figure}[H]
	\centering
	\includegraphics[width=1\linewidth]{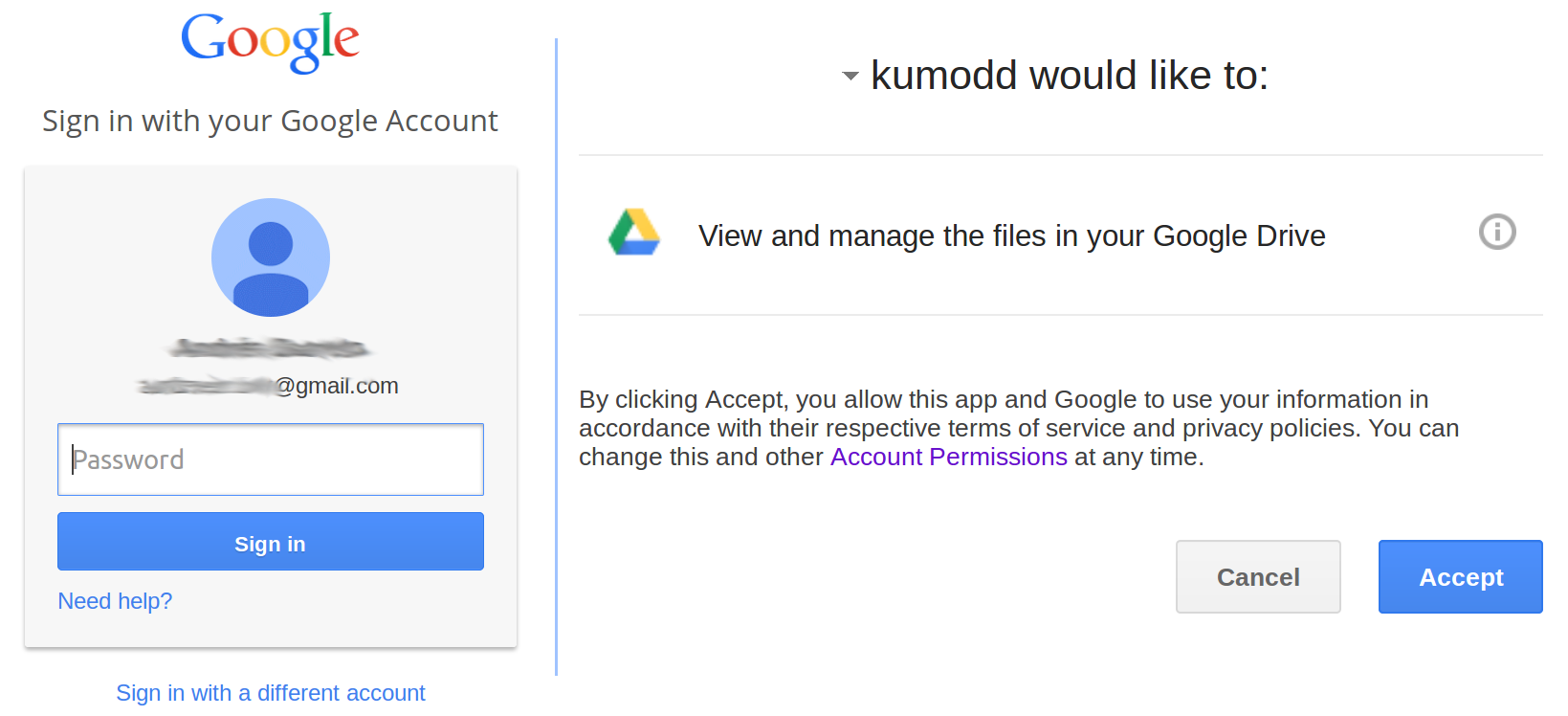}
	\caption{Authentication steps 2 and 3: provide account credentials (left) and authorize application (right)}
	\label{fig:auth23}
\end{figure}

\FloatBarrier
After supplying the correct credentials and authorizing the app (Figure~\ref{fig:auth23}), the service returns an access code, which the user has to input in the command line to finish the authentication and authorization process for the account.
If the authentication is successful, the provided access token is cached persistently in a \texttt{.dat} file is saved under \texttt{/config} folder with the name of the service. Future requests will find the token and will not prompt the user for credentials (Listing~\ref{lst:auth4}).

\begin{lstlisting}[caption=Sample processing output (with cached authorization token),label=lst:auth4]
kumo@ubuntu:~/kumodd$ python kumodd.py -s gdrive -d all
Working...
TIME(UTC) APPLICATION  USER FILE-ID REMOTE PATH REVISION LOCAL PATH HASH(MD5)
2015-06-25 03:48:43.600028 kumodd-1.0 example.dev@gmail.com 1L-7o0rgPT2f6oX60OPtF4ZUFOmOJW1Crktr3DPril8o My Drive/ppt test   v.2 downloaded/example.dev@gmail.com/My Drive/ppt test -
2015-06-25 03:48:44.951131 kumodd-1.0 example.dev@gmail.com 1huaRTOudVnLe4SPMXhMRnNQ9Y_DUr69m4TEeD5dIWuA My Drive/revision doc test   v.3 downloaded/example.dev@gmail.com/My Drive/revision doc test -
...
2015-06-25 03:48:54.254104 kumodd-1.0 example.dev@gmail.com 0B4wSliHoVUbhUHdhZlF4NlR5c3M My Drive/test folder/stuff/more stuff/tree.py   v.1 downloaded/example.dev@gmail.com/My Drive/test folder/stuff/more stuff/tree.py 61366435095ca0ca55e7192df66a0fe8
9 files downloaded and 0 updated from example.dev@gmail.com drive
Duration: 0:00:13.671442
\end{lstlisting}

\subsubsection{Content discovery}
\begin{figure}[H]
	\centering
	\includegraphics[width=1\linewidth,trim={0 19cm 0 0},clip,frame]{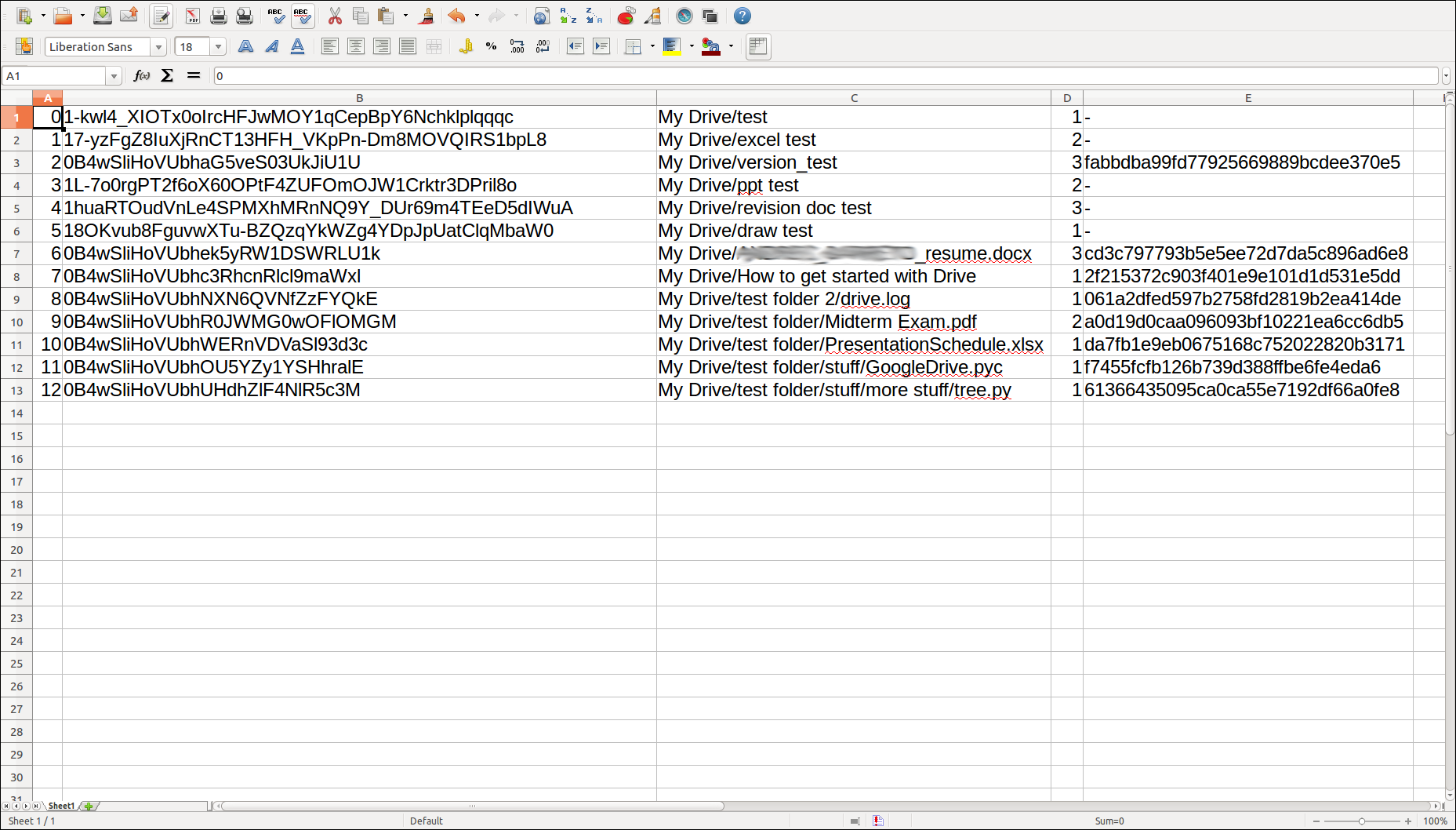}
	\caption{Contents of the generated CSV file}
	\label{fig:csv}
\end{figure}

In the current implementation, the discovery is implemented by the \emph{list} (\texttt{-l}) command, which acquires the file metadata from the drive.
As with most web services, the response is in JSON format; the amount of attribute information varies (widely) based on the provider and can be quite substantial (e.g., \gdrive).
Since it is impractical to show all of it, the \kumodd \emph{list} command outputs an abbreviated version with the most essential information formatted as a plaint text table (Listing~\ref{lst:gdrive-ls}).
The rest is logged as a CSV file in the \texttt{/localdata} folder with the name of the account and service.
The stored output can be further processed either interactively, by using a spreadsheet program (Figure~\ref{fig:csv}), or by using Unix-style command line tools, thereby enabling subsequent \emph{selective} and/or prioritized acquisition.

\begin{lstlisting}[caption=List of all files in a \gdrive account (trimmed),label={lst:gdrive-ls}]
andres@ubuntu:~/kumodd$ python kumodd.py -s gdrive -l all
Working...
FILE-ID REMOTE PATH REVISION HASH(MD5)
...1qCepBpY6Nchklplqqqc My Drive/test 1 -
...oVUbhaG5veS03UkJiU1U My Drive/version_test 3 ...bcdee370e5
...
...oVUbhUHdhZlF4NlR5c3M My Drive/test folder/stuff/more stuff/tree.py ...2df66a0fe8
\end{lstlisting}

\subsubsection{Acquisition}
As already discussed, the acquisition is performed by the \emph{download} (\texttt{-d}) command and can either be performed as a single discovery-and-acquisition step, or it can be targeted by providing a list of files (with \texttt{-csv}).

A list of successfully downloaded files is displayed with information such as download date, application version, username, file ID, remote path, download path, revisions, and cryptographic hashes. This information is also logged in the file \texttt{/downloaded/<username>/<service-name>.log}
Downloaded files are located in the \texttt{/downloaded/<username>/} directory.
The complete original metadata files with detailed information of downloaded files is stored in \texttt{/downloaded/<username>/metadata/} in JSON format. \\

\noindent\emph{Revisions}

Our tool automatically enumerates and downloads all the revisions for the files selected for acquisition;
the number of available revisions can be previewed as part of the file listing (Figure~\ref{fig:csv}, column D).
During download, the individual revisions' filenames are generated by prepending the revision timestamp to the base filename and can be viewed with the regular file browser, e.g.:

\begin{lstlisting}
(2015-02-05T08:28:26.032Z) resume.docx    8.4kB
(2015-02-08T06:31:58.971Z) resume.docx    8.8kB
\end{lstlisting}

Arguably, other naming conventions are also possible but the ultimate solution likely requires a user interface that is similar to the familiar file browser, but also understands the concept of versioning and allows the analyst to both trace the history of individual documents, and to obtain snapshots of the drive at a particular point in time. \\

\noindent\emph{Cloud-native artifacts (\gdocs)}

One new challenge presented by the cloud is the emergence of \emph{cloud-native} artifacts--data objects that have no serialized representation on the local storage, and--by extension--cannot be acquired by proxy.
\gdocs is the primary service we are concerned with in this work, however, the problem readily generalizes to many SaaS/web applications.
One of the critical differences between native applications and web apps is that the code for the latter is dynamically downloaded at run time and the persistent state of the artifacts is stored back in the cloud.
Thus, the serialized form of the data (usually in JSON) is an internal application protocol that is not readily renderable with a standalone application.

In the case of \gdocs, the local \gdrive cache contains only a link to the online location, which creates a problem for forensics.
Fortunately, the API offers the option to produce a snapshot of the document/spreadsheet/presentation in several standard formats \cite{gdrive-downloads}.
including text, PDF, and MS Office.
At present, \kumodd automatically downloads a PDF snapshot of all \gdocs encountered during acquisition.
Although this is clearly a better solution to merely cloning the link from the cache, there is still a loss of forensically-important information as the internal artifact representation contains the complete editing history of the document.
We return to this problem a little later in the presentation.

\subsection*{Web UI}
\Kumodd provides an interactive web-based GUI, which designed to be served by a lightweight local web server, which is started using the \texttt{kumodd-gui.py} module:
\begin{lstlisting}[title=Starting the web GUI]
python kumodd-gui.py
kumo@ubuntu:~/kumodd$ python kumodd-gui.py
* Running on http://127.0.0.1:5000/ (Press CTRL+C to quit)
* Restarting with stat
\end{lstlisting}
After starting the server, \kumodd's GUI becomes available at \url{http://localhost:5000} and is accessible via any web browser.
(Note: At this stage, the server should only be run locally; however, with some standard security measures, it could easily be made available remotely.)

The web module employs the same drivers used with the command line application for authenticate, discovery and acquisition.
Its main purpose is to simplify the user interactions and, for the simple case of wholesale data acquisition, the process can be accomplished in three button clicks.

\begin{figure}[H]
	\centering
	\includegraphics[width=1\linewidth,frame]{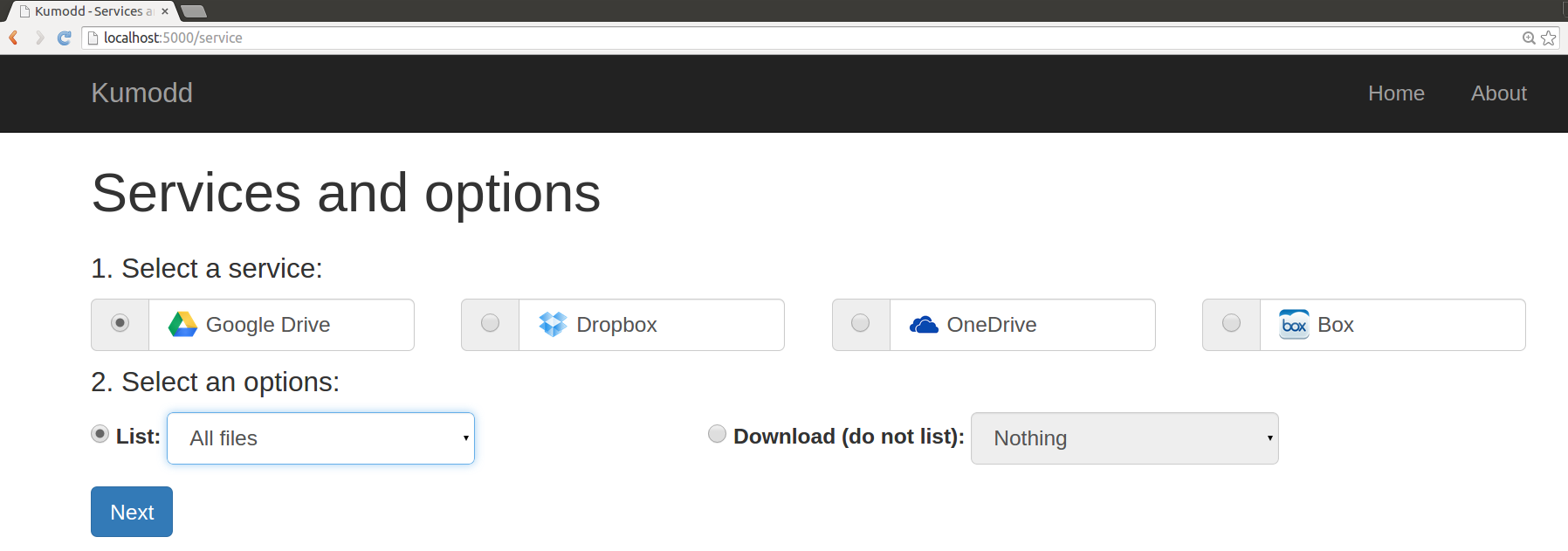}
	\caption{Web GUI: Service selection}
	\label{fig:list1}
\end{figure}

\begin{figure}[H]
	\centering
	\includegraphics[width=1\linewidth,frame]{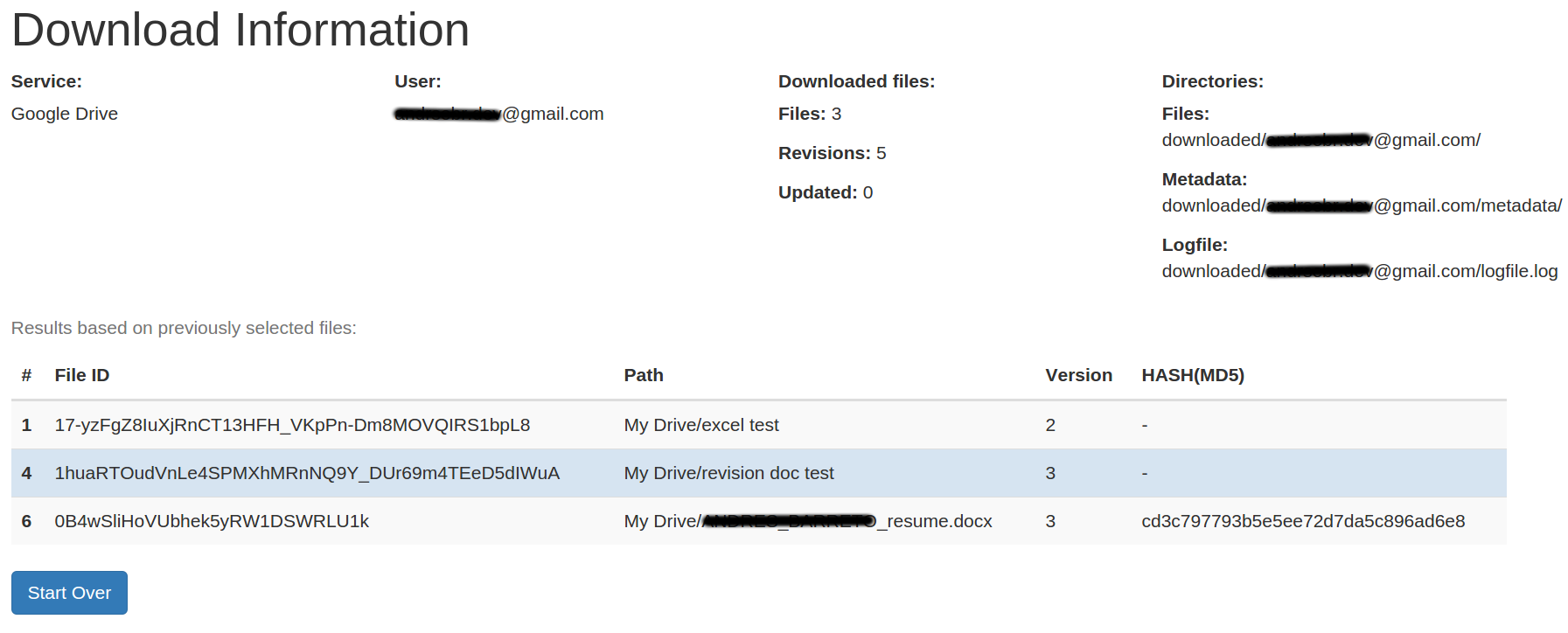}
	\caption{Web GUI: Acquisition results}
	\label{fig:list3}
\end{figure}

After pressing the \emph{Get started!} button, the user is presented with the choice of the target service and the action to perform (Figure~\ref{fig:list1}).
After this step, a detail window with a list of files and the option of choosing which to download is presented (Figure~\ref{fig:list3}).
Once the files to acquire are selected, a result screen is presented with paths to files and other relevant information.
(Every step of the process us also shown in the terminal.)

\begin{figure}[H]
	\centering
	\includegraphics[width=1\linewidth,frame]{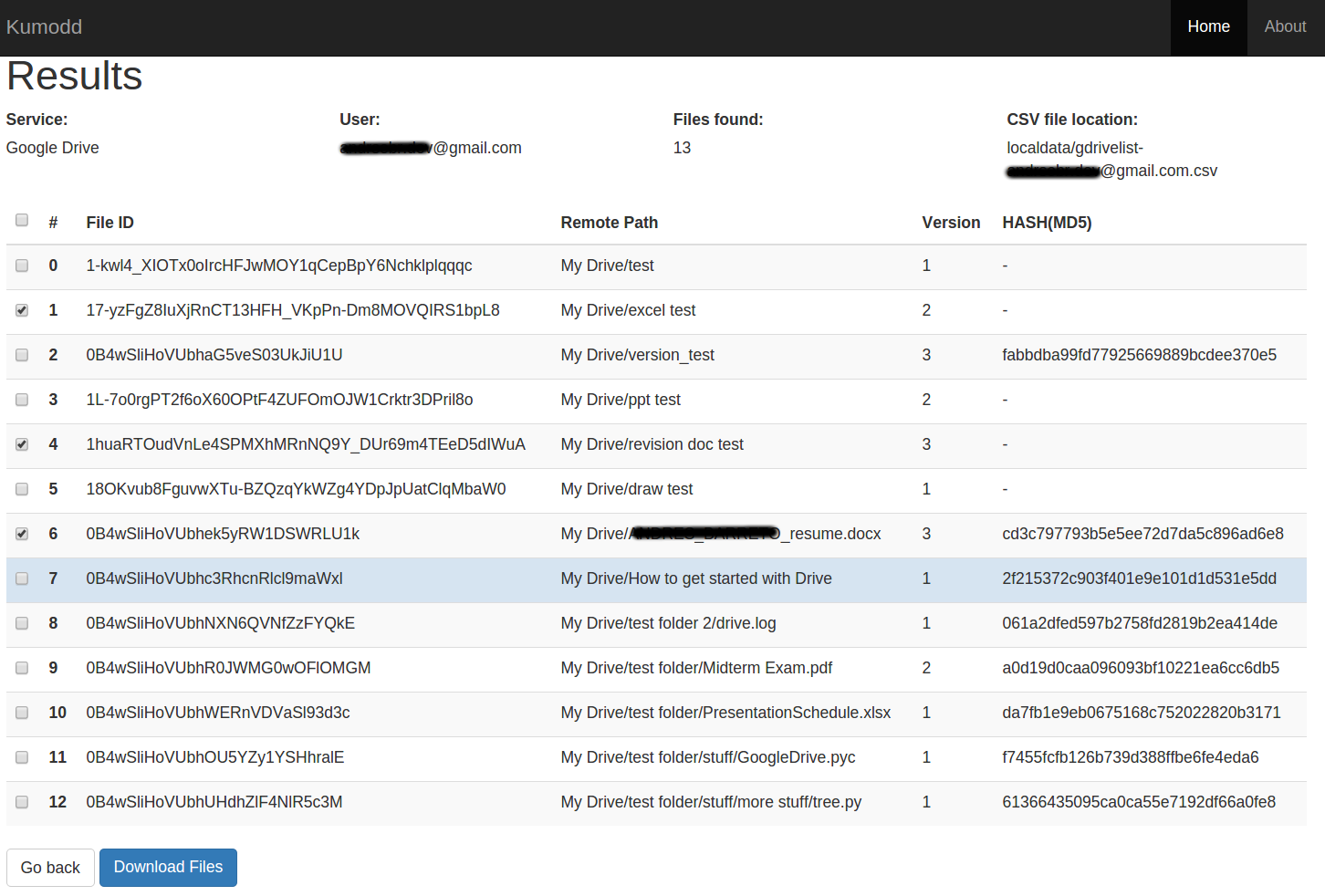}
	\caption{Web GUI: Target selection}
	\label{fig:list2}
\end{figure}

\subsection*{Validation}
To perform basic validation of the tool, we created accounts with all the services and placed known seed files. Using the normal web interface for the respective service, we created revisions and, for \gdrive, created several \gdocs artifacts. Then, we performed complete acquisition using \kumodd, and compared the cryptographic hashes of the targets. In all cases, we were able to successfully acquire all the revisions, and to obtain snapshots of the cloud-only artifacts.

\noindent\emph{Throughput}

To get a sense of the acquisition rates, we placed 1GiB of data split into 1024 files of 1MiB each; we sourced the data from the Linux pseudo-random number generator to eliminate the potential influence of behind-the-scenes compression, or de-duplication.
In preliminary tests, we used our campus network and measured the throughput at different times of day over a period of one week and did not observe any consistent correlations.

To eliminate any potential constraints spanning from our LAN, or ISP infrastructure, we moved the experiments to Amazon EC2 (US-West, N. California) instances.
For each one of the four supported services we ran complete acquisition jobs for seven days at 10:00 AM (PDT), in order to approximate daytime acquisition.
Table~\ref{tab:kumodd-times} shows the average time and throughput:

\begin{table}[h]
	\centering
	\caption{Kumodd download times and throughput}
	\begin{tabular}{c|c|c}
		Service         & Average time (mm:ss) & Throughput (MB/s) \\
		\hline\hline
		Google Drive    & 17:22     & 1.01 \\
		Dropbox         & 17:00     & 1.03 \\
		OneDrive        & 18:21     & 0.95 \\
		Box             & 18:22     & 0.95 \\
		\hline
        Average         & 17:46     & 0.98 \\
	\end{tabular}
	\label{tab:kumodd-times}
\end{table}

The results were entirely consistent with our on-campus experience, which also averaged around 1MB/s.
The most likely explanation is that the bottleneck is courtesy of cloud drive providers' bandwidth throttling.
It would appear that, at the free level, 1MB/s is the implied level of service (at least for third-party apps).

%% file: sections/05-discussion.tex
\section*{Discussion}
Before we conclude, we would like to briefly raise some of the issues that came up during our research and experimentation.

\emph{Integrity assurance}.
One of the concerning issues is the fact that quite a few services do not provide a cryptographic hash of the content of a file as part of the metadata.
For our experimental set, this includes \dbox (one of the biggest providers), which only provides a ``rev'' attribute that guaranteed to be unique, and is referred to as a ``hash'' in the API documentation\cite{dropbox-best}.
However, the generating algorithm is unknown and the observed values are way too short (they fit in 64 bits) to be of cryptographic quality.
We do know that, as of 2011, \dbox was using SHA256 on 4MiB blocks for the purposes of backend deduplication--a fact which (along with a sloppy implementation) lead to the \emph{Dropship} attack \cite{dropship}.
It is also known that the same provider uses cryptohashes to prevent the sharing of files for which it has received DMCA notice \cite{ars-dropbox-dmca}.
We would suggest that providing a cryptohash for data objects' content would be a reasonable requirement for any cloud API used for forensic purposes.

\emph{Pre-acquisition content filtering}.
One aspect that we did not address in this work is the need to provide access to the search capability built into many of the more advanced services.
This would allow triage and exploration with zero pre-processing overhead; some services also provide preview images that could also be used for this purpose.
Even with the current implementation, it is possible to filter data in/out by cryptohashes without paying the overhead of calculating them.

\emph{Multi-service management \& forward deployment}.
Cloud storage has become a \emph{de facto} commodity business with numerous providers vying for the attention of consumers and businesses;
thus, a number of cloud storage \emph{brokers} have emerged to help users manage multiple accounts and optimize their services (often at the free level).
For example, \emph{Otixo} (\url{http://otixo.com}) supports 35 different services and facilitates data movement among them.
Forensic software needs similar capabilities \emph{now} to support cloud forensics.

Looking ahead, we will soon need solutions that can be forward deployed on a cloud provider's infrastructure.
This will become necessary as cloud data grows much faster than the available bandwidth over the WAN, thereby making remote acquisitions impractical.
\Kumodd's web interface is a sketch of this type of solution; the forensic VM instance could be colocated in the same data center as the target, while the investigator controls it remotely.
In this scenario, the forensic analysis could begin immediately, while the data acquisition could be done in the background.

\emph{Long-term preservation of cloud-native artifacts}.
We already mentioned that \gdocs data objects are very different from most serialized artifacts that analysts are familiar with.
The most substantial difference is that the latter are a snapshot of the state of the artifacts, whereas the former is literally a log of user edit actions since the creation of the document.
This presents a dilemma for forensics: a) should we acquire a snapshot in a standard format, such as PDF (as \kumodd does), thereby losing all historical information; or b) acquire the \emph{changelog} and be dependent on the service provider (Google) to render it at any point in the future.
Since neither option is satifactory, we are working on a third option which will enable us to both keep the log and replay it independently of the service provider.

%% file: sections/06-conclusion.tex
\section*{Conclusion}
The main contributions of this work are:

\emph{New acquisition model}. We argued that cloud drive acquisition cannot be performed on the client in a forensically sound manner.
Specifically, the client is not guaranteed to mirror all the data in the first place, and also has no facilities to represent file revisions and the content of cloud-native artifacts.
The proper approach is to go directly to the source--the master copy maintained by the cloud service--and acquire the data via the API.
In addition to being the only means to guarantee forensically complete copy of the target data, the API approach allows for reproducibility of results, rigorous tool testing (based on well-defined API semantics), and preliminary triage of the data (via hashes and/or search APIs).
The overall development effort is significantly lower as the entire blackbox reverse-engineering aspect of client-centric approaches is eliminated.
As a reference, each of the four drivers for the individual service contains between 232 and 620 lines of Python code.

\emph{New acquisition tool}. We introduced a new tool--\kumodd--which can perform cloud drive acquisition from four major providers: \gdrive, \dbox, \boxnet, and \onedrive.
Although its primary purpose is to serve as a research platform, we expect that will quickly evolve into a reliable, open-source tool that will cover an expanding range of cloud services.

\emph{New research questions}. Based on our experience to-date, we have posed several new research problems that need to be addressed by forensic researchers.
In particular, we need to develop the means to extract, store, and replay the history of cloud-native artifacts (such as \gdocs);
we need to develop a mechanism to ensure the integrity of the data acquired from \emph{all} providers;
we need to build the tools to handle multi-service cases and forward deployment scenarios.

As a parting thought, we hope that this work will spur a different approach to cloud forensics and will serve as a cautionary note that simply extending what we know--client-side forensics--is not a promising approach.
Over the short term, this will mean expending some extra effort to develop a new toolset; over the medium-to-long term, the emphasis of logical acquisition (of which this work is an example) will allow us much greater levels of automation in the acquisition and processing of forensic targets.

%% file: ifip-kumodd.bbl
\begin{thebibliography}{10}
\providecommand{\url}[1]{#1}
\csname url@samestyle\endcsname
\providecommand{\newblock}{\relax}
\providecommand{\bibinfo}[2]{#2}
\providecommand{\BIBentrySTDinterwordspacing}{\spaceskip=0pt\relax}
\providecommand{\BIBentryALTinterwordstretchfactor}{4}
\providecommand{\BIBentryALTinterwordspacing}{\spaceskip=\fontdimen2\font plus
\BIBentryALTinterwordstretchfactor\fontdimen3\font minus
  \fontdimen4\font\relax}
\providecommand{\BIBforeignlanguage}[2]{{%
\expandafter\ifx\csname l@#1\endcsname\relax
\typeout{** WARNING: IEEEtran.bst: No hyphenation pattern has been}%
\typeout{** loaded for the language `#1'. Using the pattern for}%
\typeout{** the default language instead.}%
\else
\language=\csname l@#1\endcsname
\fi
#2}}
\providecommand{\BIBdecl}{\relax}
\BIBdecl

\bibitem{nist-cloud}
\BIBentryALTinterwordspacing
P.~Mell and T.~Grance, ``The {NIST} definition of cloud computing.'' [Online].
  Available:
  \url{http://csrc.nist.gov/publications/nistpubs/800-145/SP800-145.pdf}
\BIBentrySTDinterwordspacing

\bibitem{rightscale-2015}
\BIBentryALTinterwordspacing
Right{S}cale. (2015) Right{S}cale 2015 state of the cloud report. [Online].
  Available:
  \url{http://assets.rightscale.com/uploads/pdfs/RightScale-2015-State-of-the-Cloud-Report.pdf}
\BIBentrySTDinterwordspacing

\bibitem{gartner-2014}
\BIBentryALTinterwordspacing
Gartner. Gartner's 2014 hype cycle for emerging technologies maps the journey
  to digital business. [Online]. Available:
  \url{http://www.gartner.com/newsroom/id/2819918}
\BIBentrySTDinterwordspacing

\bibitem{gartner-hype}
\BIBentryALTinterwordspacing
------. Gartner hype cycle. [Online]. Available:
  \url{http://www.gartner.com/technology/research/methodologies/hype-cycle.jsp}
\BIBentrySTDinterwordspacing

\bibitem{article-chung}
\BIBentryALTinterwordspacing
H.~Chung, J.~Park, S.~Lee, and C.~Kang, ``Digital forensic investigation of
  cloud storage services,'' vol.~9. [Online]. Available:
  \url{http://dx.doi.org/10.1016/j.diin.2012.05.015}
\BIBentrySTDinterwordspacing

\bibitem{hale13-amazon-drive}
\BIBentryALTinterwordspacing
J.~Hale, ``Amazon cloud drive forensic analysis,'' vol.~10, pp. 295--265,
  October 2013. [Online]. Available:
  \url{http://dx.doi.org/10.1016/j.diin.2013.04.006}
\BIBentrySTDinterwordspacing

\bibitem{quick13-dropbox}
\BIBentryALTinterwordspacing
D.~Quick and K.~R. Choo, ``Dropbox analysis: Data remnants on user machines,''
  vol.~10, pp. 3--18, June 2013. [Online]. Available:
  \url{http://dx.doi.org/10.1007/978-3-642-24212-0_3}
\BIBentrySTDinterwordspacing

\bibitem{quick14-gdrive}
\BIBentryALTinterwordspacing
D.~Quick and K.-K.~R. Choo, ``Google drive: Forensic analysis of data
  remnants,'' \emph{Journal of Network and Computer Applications}, vol.~40, pp.
  179 -- 193, 2014. [Online]. Available:
  \url{http://www.sciencedirect.com/science/article/pii/S1084804513002051}
\BIBentrySTDinterwordspacing

\bibitem{martini13-owncloud}
\BIBentryALTinterwordspacing
B.~Martini and K.-K.~R. Choo, ``Cloud storage forensics: owncloud as a case
  study,'' \emph{Digital Investigation}, vol.~10, no.~4, pp. 287 -- 299, 2013.
  [Online]. Available: \url{http://dx.doi.org/10.1016/j.diin.2013.08.005}
\BIBentrySTDinterwordspacing

\bibitem{gdrive-downloads}
\BIBentryALTinterwordspacing
Google. Drive {REST} {API}--download files. [Online]. Available:
  \url{https://developers.google.com/drive/web/manage-downloads}
\BIBentrySTDinterwordspacing

\bibitem{dropbox-best}
\BIBentryALTinterwordspacing
Dropbox. Core {API} best practices. [Online]. Available:
  \url{https://www.dropbox.com/developers/core/bestpractices}
\BIBentrySTDinterwordspacing

\bibitem{dropship}
\BIBentryALTinterwordspacing
D.~DeFelippi. Dropship. [Online]. Available:
  \url{https://github.com/driverdan/dropship}
\BIBentrySTDinterwordspacing

\bibitem{ars-dropbox-dmca}
\BIBentryALTinterwordspacing
K.~Orland. Dropbox clarifies its policy on reviewing shared files for {DMCA}
  issues. [Online]. Available:
  \url{http://assets.rightscale.com/uploads/pdfs/RightScale-2015-State-of-the-Cloud-Report.pdf}
\BIBentrySTDinterwordspacing

\end{thebibliography}
